\documentclass[
reprint,
 amsmath,amssymb,
 aps,
]{revtex4-1}

\usepackage[margin=0.75in]{geometry}
\usepackage[font=footnotesize,labelfont=bf]{caption}
\usepackage[font=footnotesize,labelfont=bf]{subcaption}
\usepackage{float}
\usepackage{amsbsy}
\usepackage{amsfonts}
\usepackage{graphicx}
\usepackage{multirow}
\usepackage{mathrsfs}
\usepackage{bm}
\usepackage[hidelinks]{hyperref}
\usepackage{setspace}
\usepackage{dcolumn}

\newtheorem{theorem}{Theorem}[section]

\newtheorem{claim}[theorem]{Claim}

\begin{document}

\title{Characterising submonolayer deposition via
visibility graphs}

\author{D. Allen}
 \altaffiliation{Electronic mail: damien.allen@strath.ac.uk}
\author{M. Grinfeld}
 \email{Electronic mail: m.grinfeld@strath.ac.uk}
\affiliation{
Department of Mathematics and Statistics, University of Strathclyde, Glasgow \\
}

\author{P. A. Mulheran}
 \homepage{Electronic mail: paul.mulheran@strath.ac.uk}
\affiliation{
 Department of Chemical and Process Engineering, University of Strathclyde,
Glasgow \\
}

\date{\today}

\begin{abstract}
We use visibility graphs as a tool to analyse the results of kinetic
Monte Carlo (kMC) simulations of submonolayer deposition in a
one-dimensional point island model. We introduce an efficient
algorithm for the computation of the visibility graph resulting from a
kMC simulation and show that from the properties of the visibility
graph one can determine the critical island size, thus demonstrating
that the visibility graph approach, which implicitly combines size and
spatial data, can provide insights into island nucleation and growth
processes.

\end{abstract}

\maketitle

\section{Introduction}

Submonolayer deposition (SD) is a term used to describe the initial
stages of thin film growth, such as during molecular beam epitaxy,
where monomers are deposited onto a surface, diffuse and form
large-scale structures (islands). Of particular interest is the
mechanism for island nucleation, and how it is reflected in the
statistical properties of the growing structures.  SD is widely
studied using kinetic Monte Carlo (kMC) simulations and it is
recognised that under suitable conditions (described below), the
statistical properties of the growing structures display scale
invariance with distributions reflecting the underlying nucleation
mechanism \cite{Amar}.

Below we will be considering point islands, i.e. islands whose extent
and internal structure have been neglected.  We consider a
one-dimensional model. Both these choices have been made for
simplicity as the goal of the paper is a ``proof of concept'', to
demonstrate the ability of visibility graphs (VGs) to extract
mechanism information from kMC. That said, point islands are often used
in SD models, as they approximate SD accurately when the islands are
``well separated" \cite{Mulheran1}. In the Conclusions sections we will
discuss generalizations of our method to extended islands and to
higher-dimensional settings. Our generalisation to extended islands requires a good understanding
of the point island case and thus provides further motivation for
using point island models. 

Thus, we consider the situation where monomers are randomly
deposited onto an initially empty one-dimensional lattice $L$ at
a deposition rate of $F$ monolayers per unit time ($t$). The monomers
diffuse at a rate $D$ and islands nucleate when $i + 1$ monomers
coincide at a lattice site. We call $i$ the critical island size. We
assume no monomers can evaporate from the lattice and so the coverage
$\theta$ can be defined as $\theta = 100F t\%$.  $\theta$ is
chosen large enough for us to be in the aggregation regime (where
scale-invariance is found) i.e. where the monomers are much more
likely to aggregate into islands than nucleate into new islands. The
appropriate value of $\theta$ where the aggregation regime starts is
dependent on $i$ and the ratio $R= D/F$.

\begin{figure}[H]
\centering
\includegraphics[width=.23\textwidth]{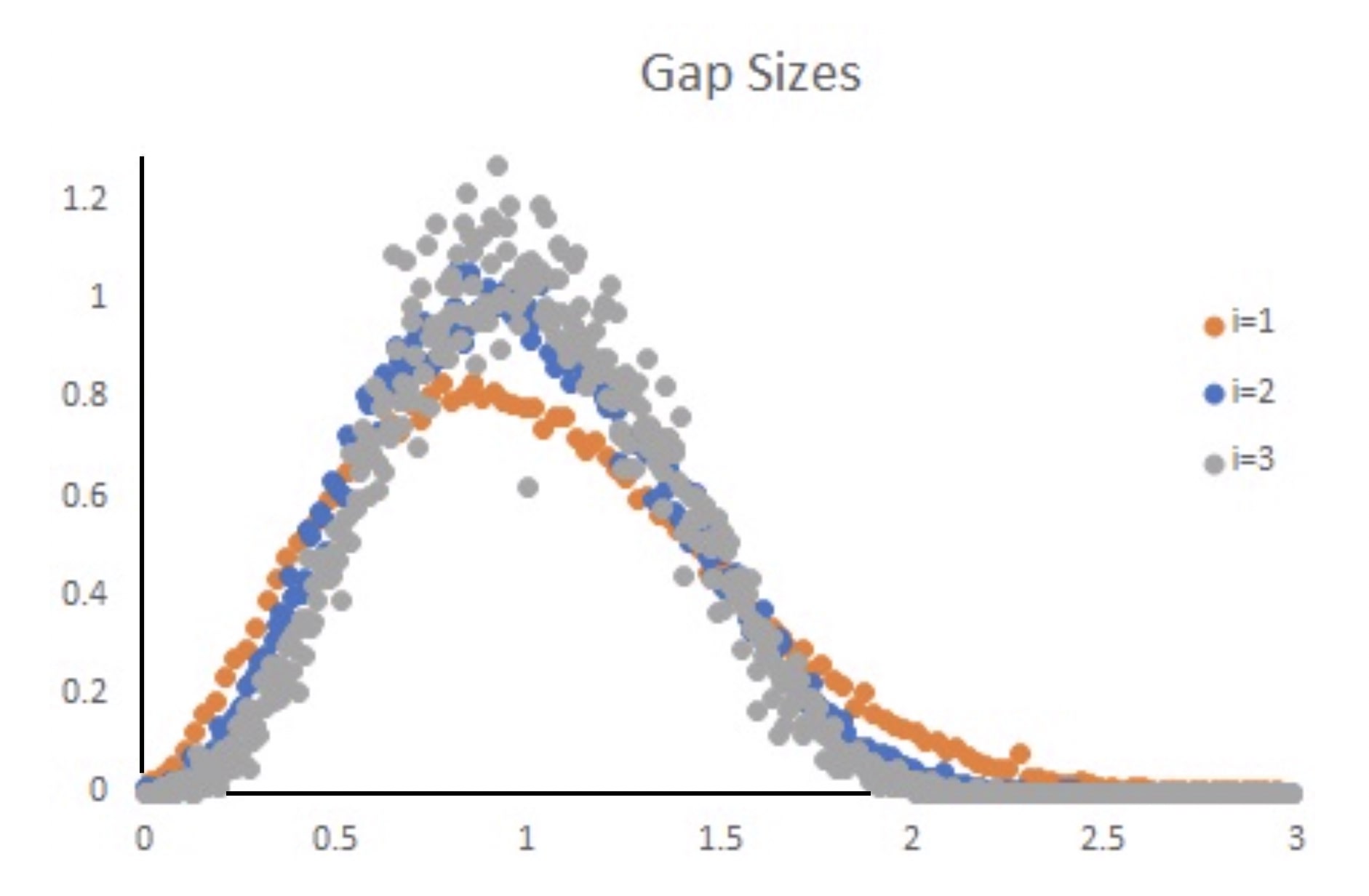}
\includegraphics[width=.23\textwidth]{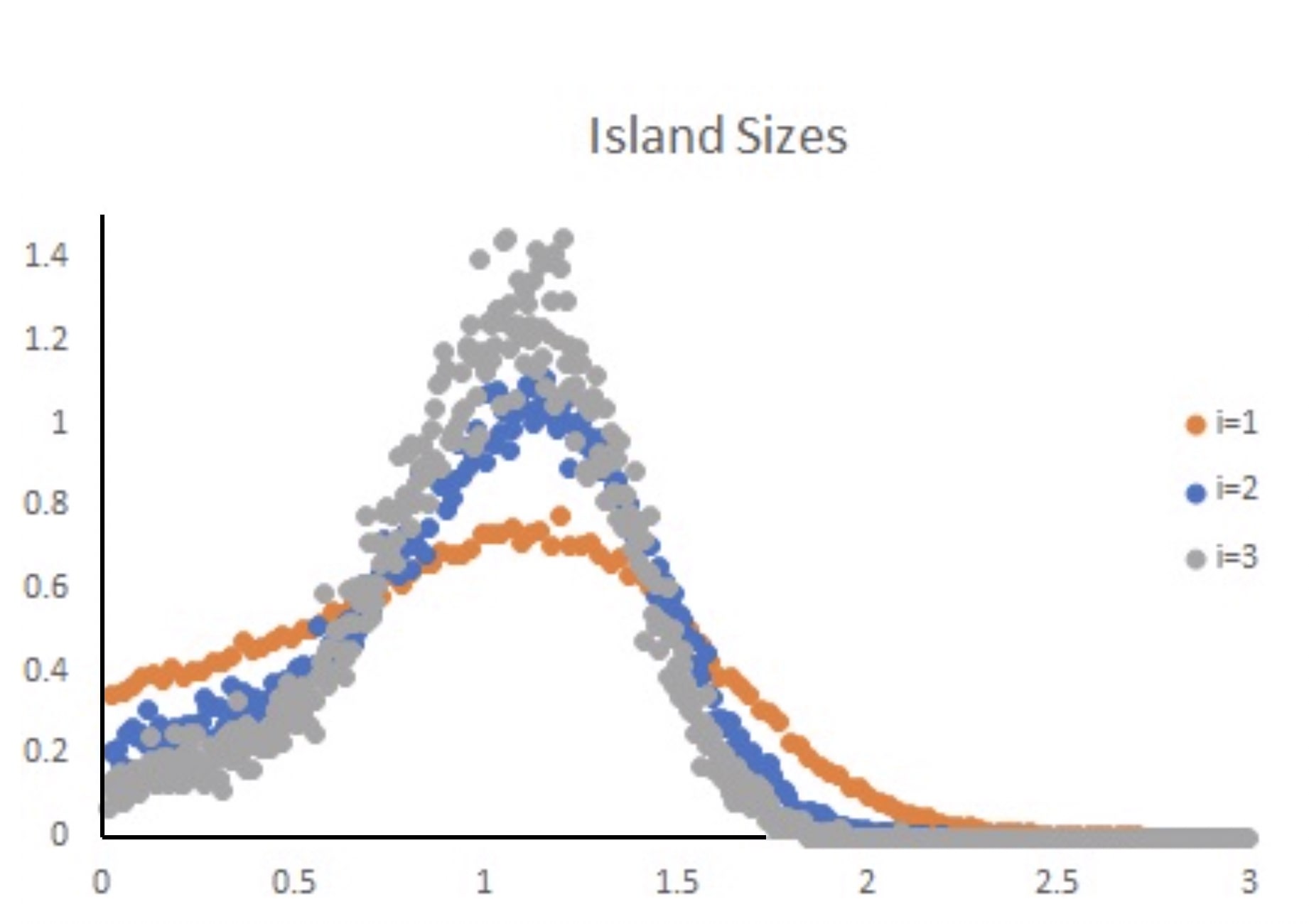}
\caption{The scaled gap and island size distributions generated from kMC simulations on lattices with $L = 10^6$
  sites, with $R= 10^6$ and up to coverage of $\theta = 200\%$, where sizes are reported relative to the average,
  averaging results over 50 runs \cite{Mulheran1}. \label{fig1} }
\end{figure}

Previous work on SD models has focused on the mean field densities,
spatial distribution of islands or the size statistics of their
islands (see \cite{Mulheran1,Mulheran2, Blackman,Einax,Evans}
for details) and the scaled gap and island size distributions
for different critical island sizes $i$ illustrated in Figure
\ref{fig1} (for limited data).  We define the {\em gap\/}
between islands as the distance between two nucleated sites and the
{\em island size\/} as the number of monomers at
a particular site on the lattice (where the size of the island must be
at least $i+1$).  It is worth noting that spatial distributions and
size statistics of islands are only negligibly affected by variations in
the number of lattice sites $L$, by variations in $R$, or by the
particular $\theta$ in the aggregation regime \cite{Blackman}.

One would like to combine the information contained in the spatial
distribution of islands and in their size statistics. A suitable tool,
which also allows higher-dimensional extensions which we discuss in
the Conclusions section, is offered by VGs introduced by
Lacasa {\em et al.}  \cite{Lacasa} originally to bring the tools of
network theory to bear on time-series analysis. Subsequently,
VGs have been used to analyse, among other things,
exchange rate series \cite{Yang} and to make solar cycle predictions
\cite{Zou}. Using VGs in the context of kMC simulations
of SD allows us to apply complex network theory to SD. Here we develop
an efficient algorithm for processing the data, and show how
properties of the VG can be used to identify the
critical island size from the island and gap size data.

  Briefly, in a VG we connect each point $P$ with coordinates
  (location, size) (the top of our grey bars in Figure 2) to all other
  points that ``are visible" from $P$ and analyse the resulting graph
  \cite{Lacasa}.

\begin{figure}[H]
    \centering
    \includegraphics[width=.5\textwidth]{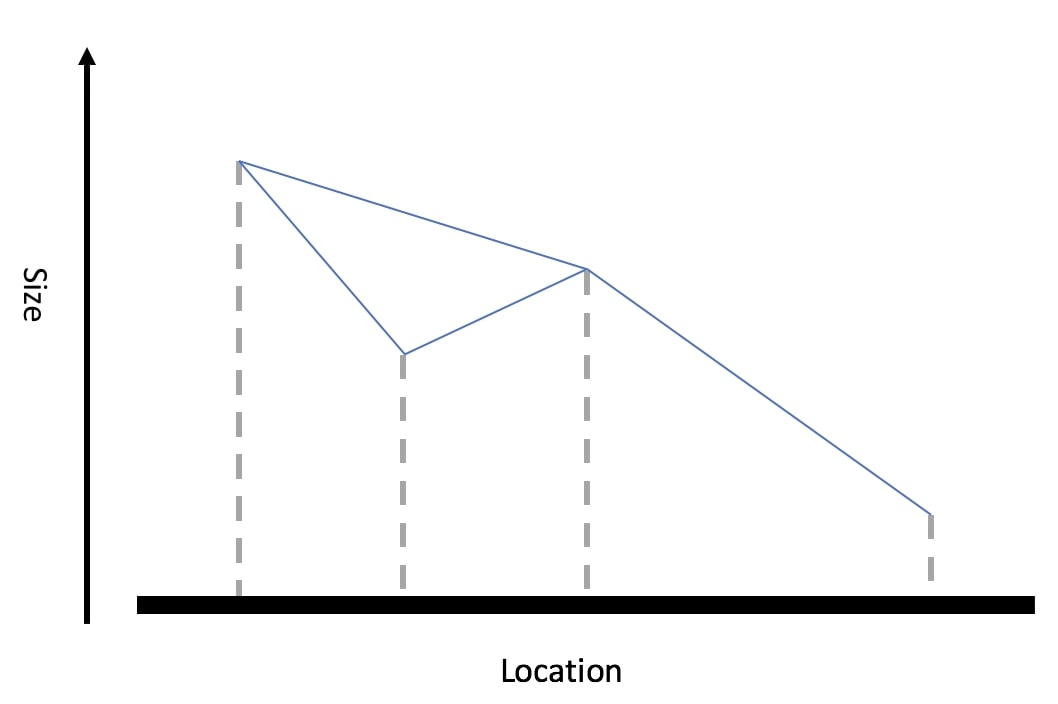}
       \caption[An example of a VG]{An example of a
         VG where the blue lines represent the edges in
         the network. \label{fig2}}
\end{figure}

\section{The VG Algorithm}

First, we would like to describe an efficient algorithm for the
computation of a VG given $n$ points in the plane, $S=\{P_1,
P_2,\ldots, P_n \}$, where $P_j = (x_j, y_j)$, $j \in \{1,\, \ldots,
n\}$.  The simplest way to construct the VG is by considering points
$P_a$, $P_b \in S$ (assuming without loss of generality that $x_a <
x_b$); then $P_a$ and $P_b$ are visible from each other if all points
$P_c \in S$ such that $x_a < x_c < x_b$, satisfy
\[
y_c < y_b+\dfrac{y_b-y_a}{x_b-x_a}(x_c-x_b).
\]

To construct the VG we need to consider all two-point
subsets of $S$, which gives us an algorithm with time complexity of
$C=\frac{1}{6}n^3+\mathcal{O}(n^2).$ As our kMC simulations produce up
to $10^5$ nucleated sites per simulation, this algorithm is
impractical as one VG takes nearly two hours to produce
on a single core desktop PC.  Hence, we aim to find an algorithm that
is faster than the na\"{i}ve one.

We collect the results needed for the construction of such an
algorithm in the following claims. Throughout, we let $P_a, P_b, P_c
\in S$ be such that $x_a<x_b<x_c$.

\begin{claim}\label{claim1}
Let $A=(a_{jk})$ where $ j,k\leq n$ be adjacency matrix of the
VG. Then $a_{jj}=0,$ $a_{jj+1} =1$ when $j< n,$
$a_{jk}=a_{kj}.$
\end{claim}

\begin{claim}\label{claim2}
Let  $P_a$ and $P_b$ be connected and  $y_a < y_b$.
Then all points $P_c$ such that $y_c < y_b$ are not visible from $P_a$.
\end{claim}

\begin{claim}\label{claim3}
Let $P_a$ be connected to $P_b$ and $P_b$ be connected to $P_c$. Then
the slopes of the line segments connecting $P_a$ to $P_b$ and $P_b$ to
$P_c$ are given by
\[
m_1=\frac{y_b-y_a}{x_b-x_a}\:\: \text{and} \:\: 
m_2=\frac{y_c-y_b}{x_c-x_b}, \: \text{respectively.}
\]
Thus
\begin{enumerate}
\item if $m_2 > m_1$, $P_c$ is visible from $P_a$,  
\item if $m_2 \leq m_1$, $P_c$ is not visible from $P_a$.
\end{enumerate}
\end{claim}

For two vectors $\bm{\mu}:=(\mu_1,\mu_2)$ and $\bm{\nu}:=(\nu_1,\nu_2)
$ we define $\bm{\mu}\wedge\bm{\nu}:=\mu_1\nu_2-\mu_2\nu_1$.

\begin{figure}[H]
    \centering
    \includegraphics[width=.4\textwidth]{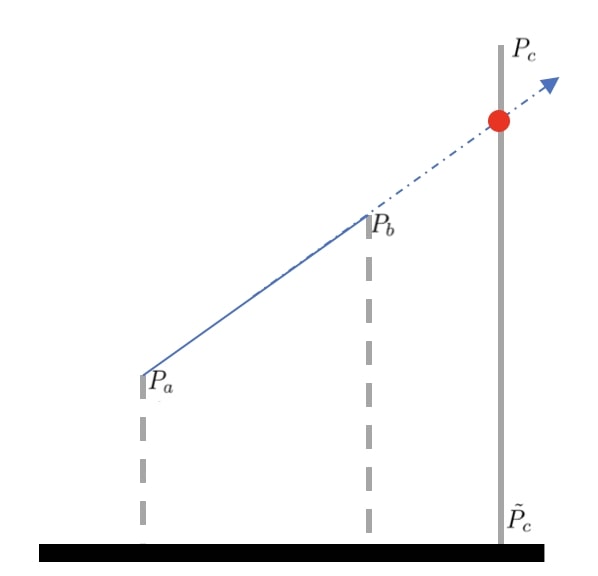}
       \caption{An illustration of \eqref{claim4} \label{fig3}}
\end{figure}

\begin{claim}\label{claim4}
Let $P_a$ be connected to $P_b$ and define
$\tilde{P_c}:=(x_c,0)$. Then $P_c$ is visible from $P_a$ if and only
if 
\[
t_1= \dfrac{| \bm{\gamma_2} \wedge \bm{\gamma_1} |}{\bm{\gamma_2}
  \cdot \bm{\gamma_3}}\in [0,\infty) \:\: \mbox{and  }
t_2= \:\: \dfrac{ \bm{\gamma_1} \cdot \bm{\gamma_3} }{\bm{\gamma_2}
  \cdot \bm{\gamma_3}} \in [0,1],
\]
where $ \bm{\gamma_1} = P_a-P_c$, $\bm{\gamma_2}=\tilde{P_c}-P_a$, and 
$\bm{\gamma_3} =-(P_b(2)-P_a(2),P_b(1)-P_a(1))$, see Figure \ref{fig3}.
\end{claim}

\subsection{New VG Algorithm }

We can now use Claims \eqref{claim1}-\eqref{claim4} to construct our
VG algorithm. We consider an arbitrary point $P_j \in S$
and the vector of elements to the right of the main diagonal in the
$j$-th row of the adjacency matrix of the VG
$[a_{j,j+1}, \, a_{j,j+2}, \ldots,\, a_{j,n}]$. Letting $2\leq k \leq
n-j$ we have:
\begin{itemize}
\item $a_{j,j+1}=1$, by Claim \eqref{claim1}.

\item If $ a_{j,j+k-1} = 1$ and $y_{j+i} < y_{j+k-1}$ where $k\leq i \leq n-j$,
then $a_{j,j+i}=0$ by Claim  \eqref{claim2}.
    
\item If $a_{j,j+k-1} = 1$ then $a_{j,j+k}=1$ if $m_2>m_1$ and
  $a_{j,j+k}=0$ otherwise by Claim  \eqref{claim3}.

\item If $a_{j,j+k-1} = 0$, then $a_{j,j+k}=1$ if $t_1 \in [0,\infty)$
  and $t_2 \in [0,1]$ and $a_{j,j+k}=0$ otherwise by Claim  \eqref{claim4}.

\end{itemize}
We continue this process for all $P_j \in S$  and then use the final
property from Claim  \eqref{claim1} to complete our adjacency matrix.

Our new algorithm is around 15 times faster than the original and
typically reduces computation time from two hours to eight minutes for
the construction of a VG from one kMC run.

\section{Characterising the VG}

Thus, we start with a kMC simulation of SD in one space
dimension. Once the simulation is complete, we mark the location and
the size (`height') of each nucleated island and construct the
resulting VG.

Our simulations were performed on lattices with $L=10^6$ sites,
$R=10^6$ up to coverage of $\theta=200\%$ for different critical
island sizes $i$. (For $i = 0$ we set the spontaneous nucleation
probability i.e. the chance a monomer becomes fixed to the lattice to
$p = 10^{-6}$). We choose these conditions to guarantee we are in the
aggregation regime and throughout the remainder of this paper we refer
to these conditions as our `standard conditions'.

There are many ways to characterise a graph; these include criteria
based on vertex degree, spectrum of the adjacency and other matrices
defined from the graph, communicability and centrality indices
\cite{Mieghem}. Below we only analyse the vertex degree distribution
and spectral gap in the adjacency matrix as these are sufficient to
differentiate between VGs corresponding to different
critical island sizes $i$. We discuss other possibilities of the
method in the Conclusions section.

\subsection{Degree Distribution}

\begin{figure}[H]
\centering
\includegraphics[width=.4\textwidth]{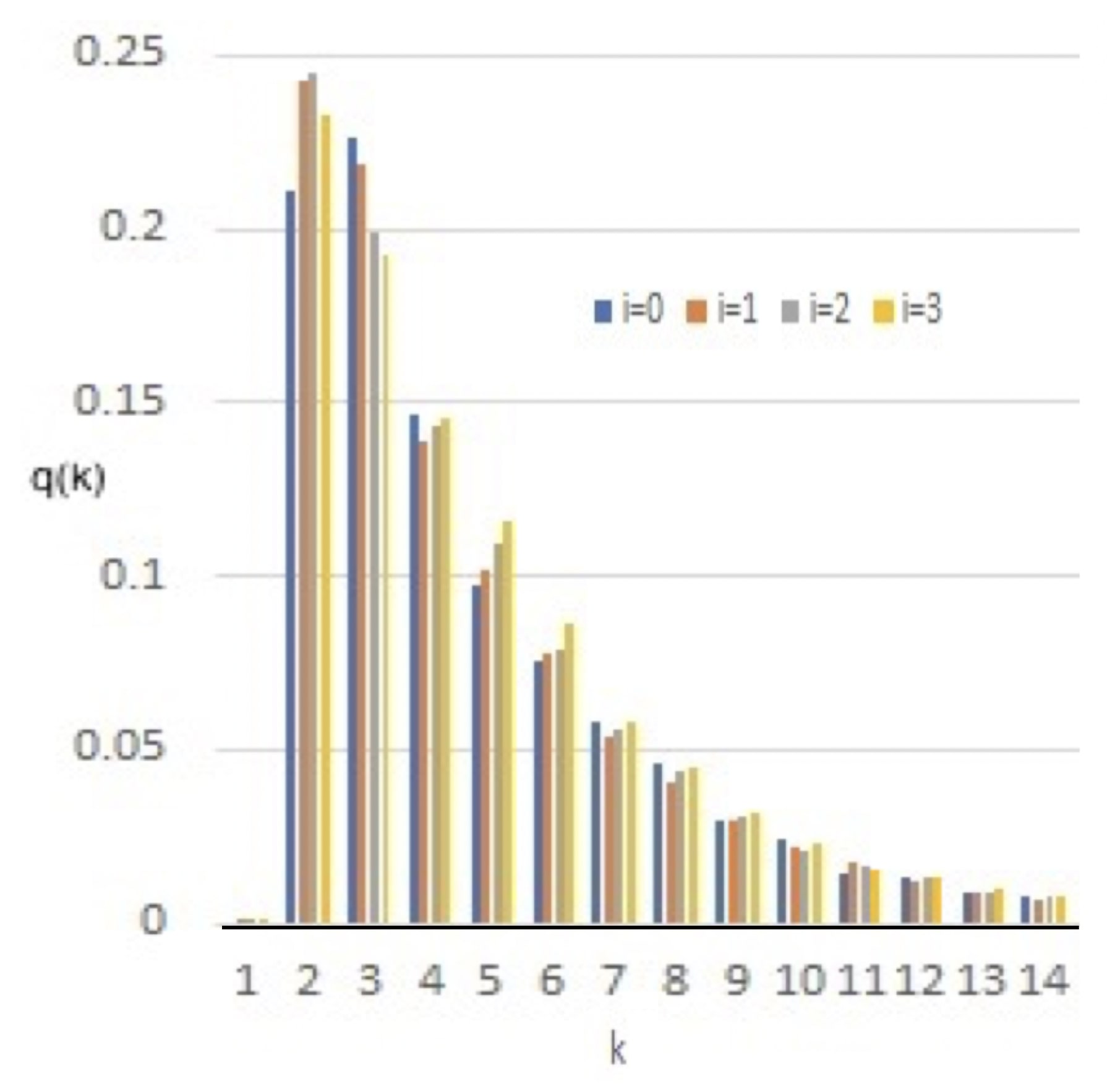}
\caption{The vertex degree distributions of VGs generated from
  kMC simulations on lattices with $L = 10^6$
  sites, with $R= 10^6$ and up to coverage of $\theta = 200\%$ 
  for $i=0,1,2$ and $3$ and in the $i=0$ case we let $p=10^{-6}$,
  averaging results over 50 runs.}\label{fig4}
\end{figure}

We begin our characterisation of the VG by considering
the vertex degree distribution. Let $n$ be the number of nodes in our
VG and $m(k)$ be the number of nodes in our visibility
graph with $k$ connectivity; for simplicity we define
$q(k) := m(k)/n$. The vertex degree distributions of VGs
generated from kMC simulations (under our standard conditions)
averaged over 50 simulations are shown in Figure \ref{fig4}.

\begin{figure}[H]
\centering
\includegraphics[width=.4\textwidth]{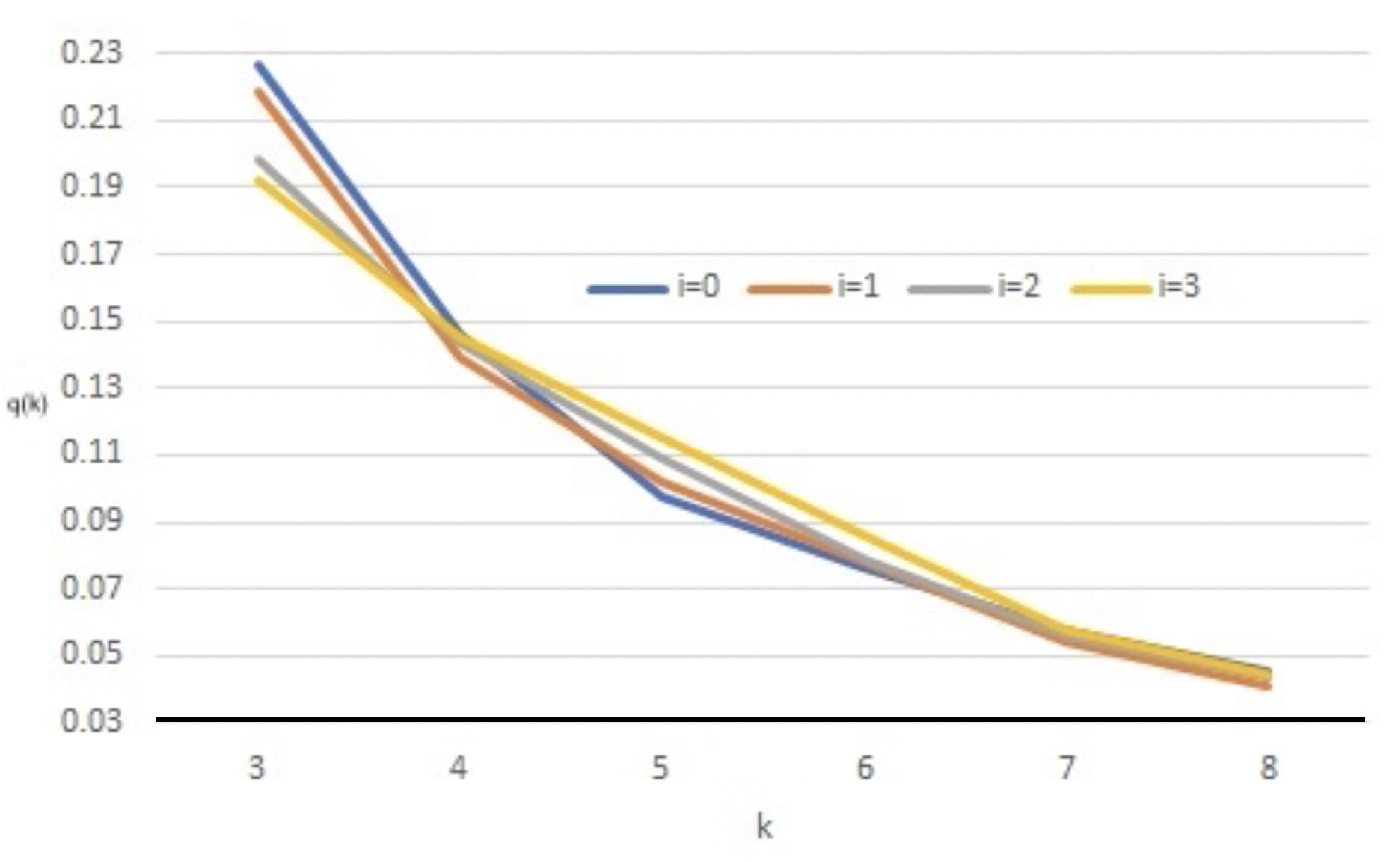}
\caption{The vertex degree distributions of VGs
  generated from kMC simulations on lattices with $L = 10^6$ sites,
  with $R= 10^6$ and up to coverage of $\theta = 200\%$ when $i=0,1,2$
  and $3$ and in the $i=0$ case we let $p=10^{-6}$, averaging results
  over 50 runs, for $3 \leq k \leq 8$.} \label{fig5}
\end{figure}

From Figure \ref{fig4}, we see that graphs corresponding to different
values of $i$ differ in the statistics of nodes having degree $k$,
particularly when $ 3 \leq k \leq 8$. To investigate this finding
further, we consider this specific region as shown in Figure
\ref{fig5}.  To emphasise the differences we connect the points with
straight lines. 

As expected, for every $i$ considered, the degree distributions are
exponential (see \cite{Allen} for further details), however there are
noticeable differences particularly for $q(3)$ for different $i$.
Changes in $R$ (when $R=10^7, 10^8$ and $10^9$), $L$ (when $L=10^7$
and $10^8$) and $\theta$ (when $\theta=100\%$) have a negligible
effect on the degree distributions confirming that we are operating in
the aggregation (scaling) regime.  This is consistent with the work on
gap size, island size and spatial distributions, see \cite{Mulheran1}.

It is interesting to see whether the values for $q(3)$ can be used to
{\em identify} the value of $i$. To do this, we
generate a VG from a set of kMC simulations and generate
$q(3)$ in each case. Our simulations were performed for values of
$i=0,1,2,$ and $3$. We performed the process $50$ times, in each
case testing whether $q(3)$ values alone can determine the value of i
used to generate the kMC data. We found that $i$ was correctly
predicted in $92\%$ of cases. In addition, in all cases the predicted
$i$ was within $1$ of the true value of $i$.

\subsection{Spectrum of the adjacency matrix}

\begin{figure}[H]
\centering
\includegraphics[width=.4\textwidth]{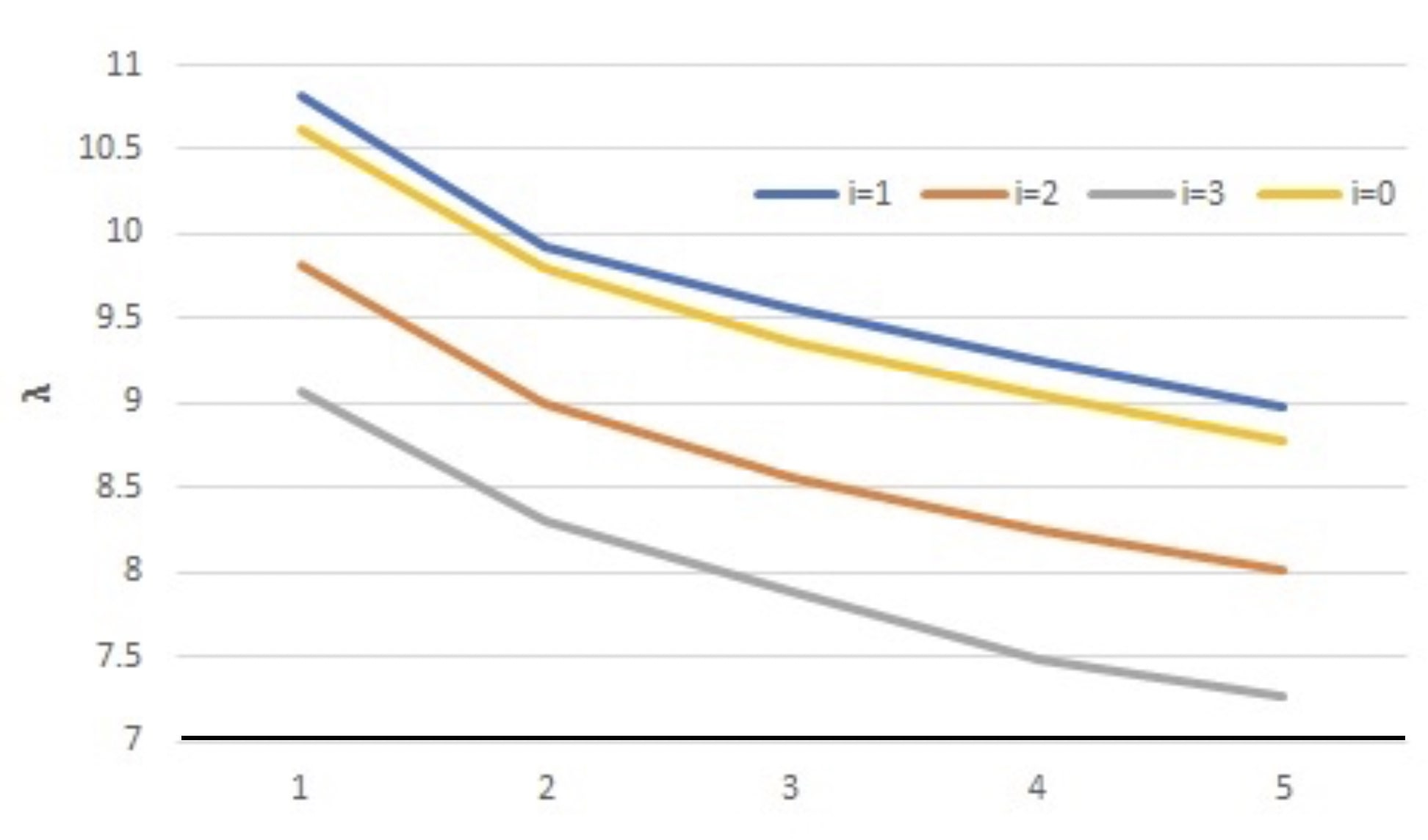}
\caption[]{ The eigenvalues of VGs generated from kMC
  simulations on lattices with $L = 10^6$
  sites, with $R= 10^6$ and up to coverage of $\theta = 200\%$ 
  when $i=0,1,2$ and $3$ and in the $i=0$ case we let $p=10^{-6}$,
  averaging results over 50 runs.}\label{fig6}
\end{figure}

Next we consider the adjacency matrix of the VG. We consider the first
five eigenvalues of the adjacency matrix of our VGs generated from kMC
simulations under our standard conditions. As with the vertex degree
distribution, we find consistent behaviour for each $i$. We average
the eigenvalues over 50 runs, as shown in Figure \ref{fig6}.

\begin{figure}[H]
\centering
\includegraphics[width=.4\textwidth]{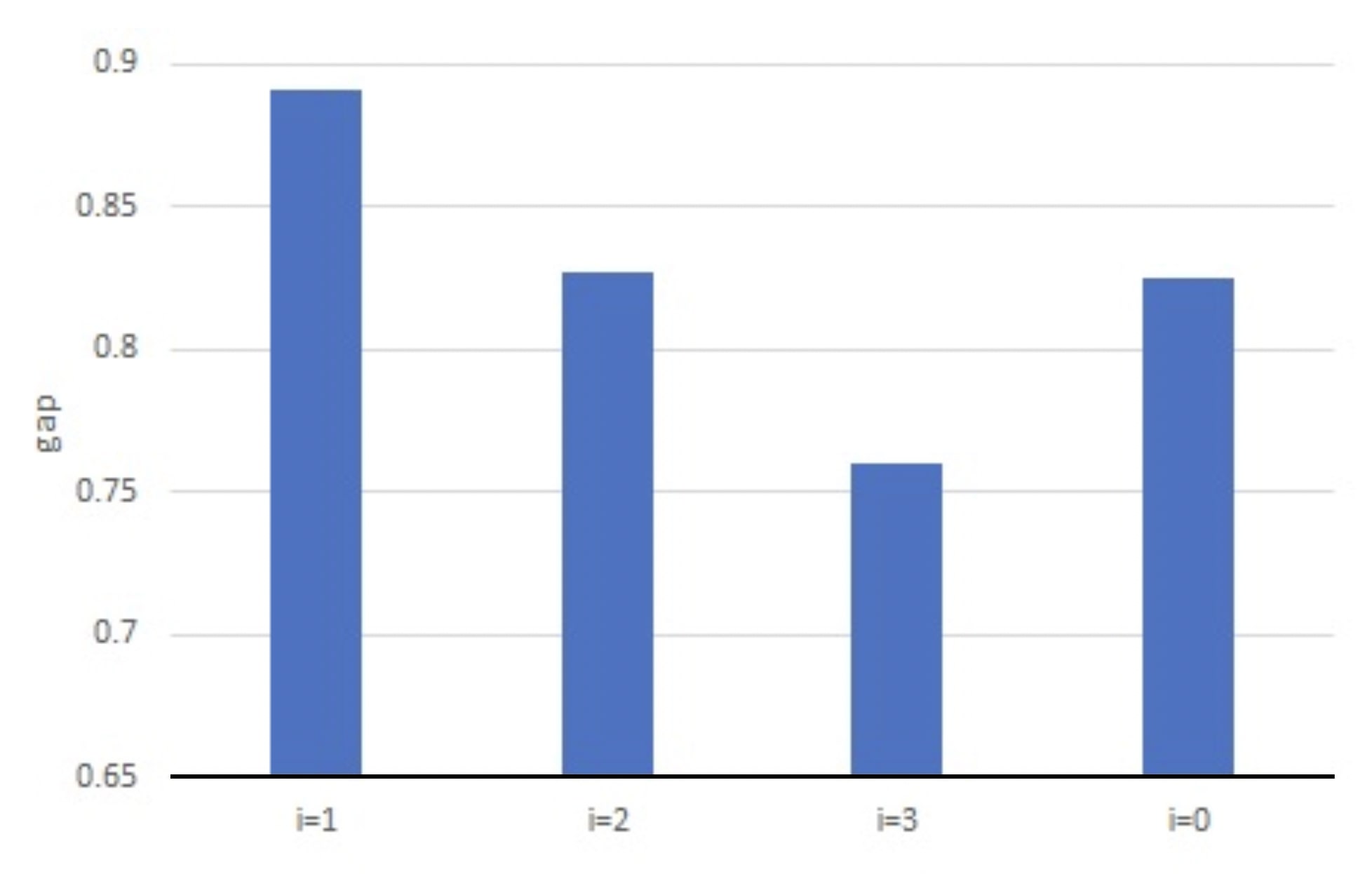}
\caption[]{ The gap between the first and second eigenvalue of the
  adjacency matrix from VGs generated from kMC
  simulations when $i=0,1,2$ and $3$, $R=16 \times 10^6$, $L=10^6$,
  $\theta=200\%$ and in the $i=0$ we let $p=10^{-6}$, averaging
  results over 50 runs.}\label{fig7}
\end{figure}

As the $i = 0$ case is practically indistinguishable from the $i = 1$
case, to separate the two we consider the gap between the largest
eigenvalue and the second largest eigenvalue of the
adjacency matrix, i.e., the spectral gap, which
has been shown to be related to the connectivity of the graph
\cite{Mieghem}; these results are shown in Figure \ref{fig7}.

Once again, we find that changes in $R$, $L$ and $\theta$ have a
negligible effect on the eigenvalues and the gap between the
eigenvalues.

As in the case of using $q(3)$ to distinguish between nucleation
mechanisms, we generate a VG from a kMC simulation under
our operational conditions averaging results over 50 runs. In order to
identify the value of $i$ from the adjacency matrix, we use the
largest eigenvalue to separate the case of $i=2,3$ from
the rest, and then the spectral gap to differentiate between $i=0$ and
$i=1$. We performed this process 50 times. We found that $i$ was
correctly predicted in all cases. Note, in contrast to Figure \ref{fig1}, the excellent separation of the $i=2$ and $i=3$ case.

\section{Conclusions}

We have shown that the analysis of some of the properties of the VG
generated from a kMC simulation allows us to determine the underlying
nucleation mechanism; both the degree distribution $(q(3))$ and the
spectrum of the adjacency matrix allow us reliably to identify the
value of $i$ used in the kMC simulation. We have also created an
efficient algorithm for processing the kMC position/size
data. Therefore, we have created an effective characterisation process
that can be applied to experimental data for SD in one dimension, such
as island nucleation and growth on a stepped substrate
\cite{Pownall}. This approach provides a way for the molecular scale
rules for nucleation and growth to be decided.

The VG method has the potential to deal with more complicated
mechanisms e.g. evaporation, mobile islands, unstable islands
\cite{Allen1}, electric fields and any level of coverage within the
scaling regime as discussed above. The generalisation of our work to
extended islands is also straightforward, as we can create the vectors
$P$ used in the construction of VG, by using the position of the
centre of mass of an island and its mass as coordinates.  We leave
these versions of SD to future work.

It is true that at this stage there is no {\em a priori\/} reason why
information about the critical island size $i$ should be contained in
$q(3)$ or in the spectrum of the adjacency matrix, as demonstrated
here. In general, assigning meaning to the spectrum of the adjacency 
matrix of a graph is difficult, as many different properties of the graph 
are stored in a single number, an eigenvalue. For a discussion of these 
issues, see \cite{Zenil}. The VG framework used here falls in the domain
of ``equation-free'' approaches (for a general philosophy of which see
\cite{Kevrekidis}), as do the applications in complex (in particular,
biological and financial) systems of topological data analysis
\cite{Carlsson} and Minkowski functionals \cite{Mecke}. Such an
exploratory study is necessary to verify, as we do here, that the tool
is up to the task.

An important question is how to extend this methodology to two and
three-space dimensions. In \cite{Lacasa2} a method is proposed to
extend one-dimensional VGs to higher dimensions which enables the
construction of VGs of large-scale spatially-extended surfaces. The
method uses one-dimensional VGs along different straight lines in the
multi-dimensional lattice to construct a single VG (only dependent on
  the number of lines one considers).

Of course, other ways of correctly identifying $i$ from data, such as
from the scaled distribution of island sizes, already exist, and it is
not clear whether a VG offers any immediate advantages
in terms of robustness against noise or clarity of interpretation when
the growth rules evolve over time. Nevertheless, we have successfully
demonstrated that the VG approach usefully combines spatial and size
data in a physically meaningful way, relating SD to network theory,
thereby opening up new approaches to understanding more complex
SD processes and their classification.

\end{document}